\documentclass[11pt]{article}
\usepackage{osid}
\usepackage{amsfonts}
\usepackage{amssymb}
\usepackage{bm}
\usepackage{graphicx}
\usepackage{amsmath}
\usepackage{float}
\usepackage{color}

\title{Single-photon single-ion interaction in free space
  configuration in front of a parabolic mirror}

\author{Magdalena Stobi\'nska \\{\footnotesize\it Institute of Theoretical Physics and Astrophysics, University of Gdansk, ul. Wita Stwosza 57, 80-952 Gdansk, Poland}\\{\footnotesize\it Institute of Physics, Polish Academy of Sciences, Al.\ Lotnik\'ow 32/46, 02-668 Warsaw, Poland}\\{\footnotesize\it \& magdalena.stobinska@gmail.com}\\[2ex]
        Robert Alicki\\{\footnotesize\it Institute of Theoretical Physics and Astrophysics, University of Gda\'nsk, ul.~Wita Stwosza 57, 80-952 Gda\'nsk, Poland} }

\begin{document}

\maketitle 

\begin{abstract}
  The efficient interaction between single photons and single matter
  objects in free space is of key importance for quantum technologies.
  An experimental setup for testing this possibility involves single
  two-level ion trapped at the focus of a parabolic metallic mirror.
  We study the conditions for the setup, under which the assumption
  about the free-space mode structure of the radiation field in the
  vicinity of the atom is justified.  In our analysis we apply
  vectorial properties of light by including polarization degree of
  freedom.  We look for possible changes in the spontaneous emission
  rate of the atom resulting from the presence of the parabolic
  boundary conditions.
\end{abstract}

\section{Introduction}

The efficient interaction between single photons and single matter
objects is of key importance for quantum technologies.  Although,
quantum electrodynamics is one of the best physical theories, this
interaction remains an intriguing research from the fundamental point
of view.

The strong matter-light coupling has applications in many branches of
physics and technology: in quantum communication for quantum repeaters
(coupling between the flying and stationary qubit), quantum computing,
distributed networks \cite{Cirac1997,Duan2001,Maitre1997} and
microscopy \cite{Hood2000,Pinkse2000,Lindlein2007}.  It is achieved in
cavity QED, where light is reflected from its mirrors and interacts
with an object inside the cavity for a long time.  The cavity supports
only one (or very few) radiation mode and in this way prevents from
unwanted interaction with other modes, which makes the evolution of
the object difficult to invert.  However, technological solutions
based on cavities QED are not scalable.  Strong atom-light interaction
in free space configuration
\cite{Gerhardt,Sandoghdar,Tey,Stobinska2009} i.e. without a cavity,
may provide us with less technologically demanding solutions for large
distances.

Possibility of such near perfect coupling in free space has already
been theoretically predicted \cite{Stobinska2009}.  It can be achieved
if the photon is prepared in a wavepacket with a suitable
spatio-temporal profile.  An experimental setup for testing this
theory is currently being built.  It involves single two-level ion
trapped at the focus of a parabolic metallic mirror (see
Fig.~\ref{fig1}) \cite{Sondermann2007}.  The mirror with an additional
electrode constitutes a Paul trap.  Its parabolic shape allows to use
it as an electric field mode converter, useful for shaping the photon
wavepackets.  However, numerous discussions arose in scientific
community whether the mirror, being a half-cavity, indeed ensures a
free space configuration.  Similar work considering effects of the
large cavity limit \cite{Alber1992} and structure of standing light
waves in half-cavity arrangement on atom decay rate was theoretically
investigated within the scalar light model
\cite{Drexhage1970,Dorner2002}.  The change of density of modes near
the atom in front of a planar mirror was experimentally verified
\cite{Eschner2001}.

In this paper we study the conditions for the setup in
Fig.~\ref{fig1}, under which the assumption about the free-space mode
structure of the electromagnetic field in the vicinity of the atom is
justified.  In our analysis we apply vectorial properties of light by
including polarization degree of freedom.  We look for possible
changes in the spontaneous emission rate of the atom resulting from
the presence of the parabolic boundary conditions.

We emphasis that the setup under discussion is especially useful for
quantum communication applications, since it allows for error
correction schemes: the shape of the mirror ensures the access to all
environmental degrees of freedom.  This distinguishing feature of the
setup makes also possible investigation of the universal model of
spontaneous emission process in free space from the fundamental point
of view.  It is irreversible for the atomic subsystem alone, but
unitary for the system as a whole \cite{Alicki2008}.

This paper is organized as follows.  In section
\ref{sec:parabolic_geometry} we discuss the setup and develop the
formalism of the normal modes genuine to the parabolic geometry,
without the mirror. In section \ref{sec:mirror} we analyze the
correction to the spontaneous emission rate resulting from presence of
the parabolic metallic mirror.  We finish the paper with the
conclusions.

\section{Decay rate in parabolic geometry}
\label{sec:parabolic_geometry}

\begin{figure}
\begin{center}
      \scalebox{0.4}{\includegraphics{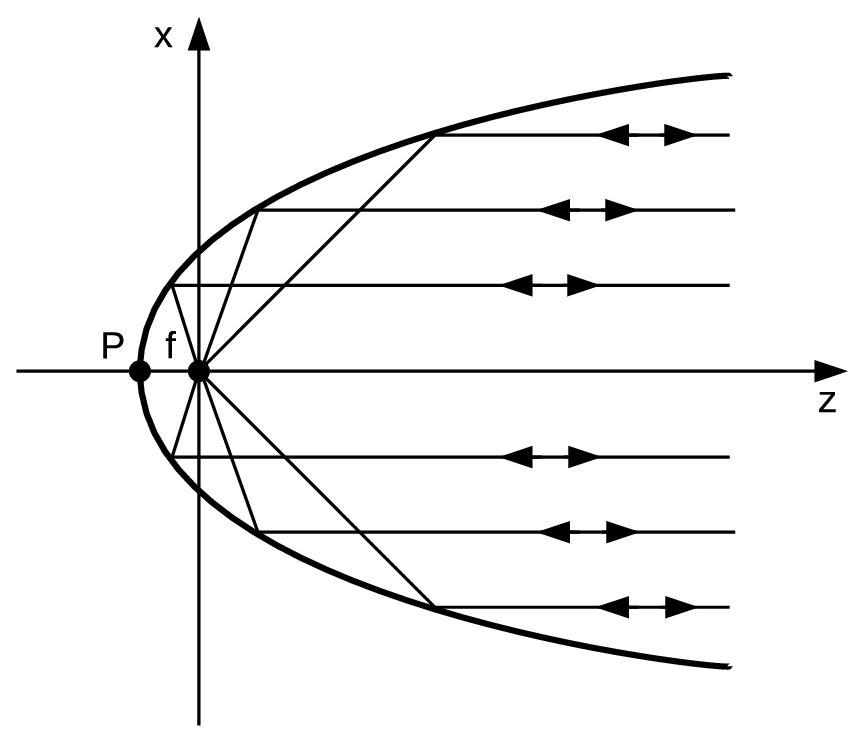}}
  \end{center}
  \caption{The experimental setup under consideration: a single
    two-level ion is trapped at the focus of a parabolic metallic
    mirror. The focal length equals $f$. The ion interacts with the
    light coming from the whole $4 \pi$ solid angle.}
  \label{fig1}
\end{figure}

Let us start the discussion with the description of the parabolic
mirror ion trap depicted in Fig.~\ref{fig1}. The ion is located in
half-open space, or in a half-cavity.  The parabolic shape of the
mirror ensures that if a light beam is sent parallel to the mirror
axis towards the ion, it interacts with the light coming from the
whole $4 \pi$ solid angle, and similarly it allows to collect the
whole light resulting from its spontaneous decay. Thinking in terms of
the classical ray-picture, the incident angle for either incoming or
outcoming light is never equal to $\pi/2$ (except from one
direction). It means that if the atom emits a photon in the
spontaneous emission process, contrary to the standard cavity case,
the radiation is never back reflected to the atom, and thus the atom
does not feel the presence of the boundary conditions, just like in
free space. Nevertheless, the mirror creates the nodes and anti-nodes
in the reflected modes and thus is able to change the electromagnetic
vacuum structure.  Therefore, in the limit where the focal length $f$
of the mirror is comparable to the wavelength $\lambda$ of the mode
resonant with the atomic transition, similarly to a small cavity or in
front of a planar mirror \cite{Morawitz1969}, the changes of the
spontaneous decay rate should be significant. We will show that
depending on the characteristic parameters of the setup: the focal
length $f$, the wavelength $\lambda$ and the orientation of the atomic
electric dipole moment, the setup provides us with either a free-space
configuration or a tailored electromagnetic reservoir near the atom.

We begin with introducing our formalism of the normal modes genuine to
the parabolic geometry without the mirror.  Since the ion is located
in a half-open space we work within the framework of the
Weisskopf-Wigner model of interaction between a single matter qubit
and a quantized radiation field.  We begin with the expansion of the
electric field operator for the radiation field in free space
using the basis (modes) suitable for the parabolic symmetry of the
problem. We follow the results of \cite{BKM} and use the modes given
by the formula
\begin{equation}
\vec{E}^{\sigma}_{k,\ell,\mu}(\vec{r})= \frac{k}{(2\pi)^{3/2}}
\int_{S^2} d\vec{n} \, e^{i k\vec{n} \cdot \vec{r}} \,
    {h}_{\ell,\mu}(\vec{n})\vec{e}^{\,\sigma}(\vec{n}),
\label{solution}
\end{equation}
where $\vec{k} = k\vec{n}$ denotes the wavevector, $\sigma = 1,2$
enumerates polarization states, parameters $\ell = 0, \pm 1, \pm 2,
...$ and $\mu\in(-\infty, +\infty)$ are the mode numbers. The unit
vector $\vec{n}$ and the polarization vectors $\vec{e}^{\,1}(\vec{n}),
\vec{e}^{\,2}(\vec{n})$ constitute the orthonormal basis what ensures
the transversality condition
\begin{equation}
\nabla \cdot \vec{E}=0.
\label{trans}
\end{equation}
We choose
\begin{eqnarray}
\vec{n} &=& (\sin\theta \cos\varphi, \sin\theta \sin\varphi,
\cos\theta),\\ \vec{e}^{\,1}(\vec{n}) &=& (\sin\varphi,-\cos\varphi,0),
\\ \vec{e}^{\,2}(\vec{n}) &=& (\cos\theta \cos\varphi, \cos\theta
\sin\varphi, -\sin\theta).
\end{eqnarray}
The manifest form of the modes ${h}_{\ell,\mu}(\vec{n})$ reads
\cite{BKM}
\begin{equation}
h_{\ell,\mu}(\theta, \varphi) =
\chi_{\mu}(\theta)\frac{e^{i\ell\varphi} }{\sqrt{ 2\pi}}
\end{equation}
where
\begin{eqnarray}
\chi_{\mu}(\theta) &=& \frac{\exp\left( -i \mu \ln[\tan \theta/2]
  \right)}{\sqrt{2\pi}\sin\theta}.
\label{solution-free}
\end{eqnarray}
One can easily check the orthogonality and completeness conditions
\begin{equation}
\int_0^{2\pi} \!\!\!\!d \varphi
\!\int_0^{\pi}\!\!\!d\theta\sin\theta\; h^*_{\ell,\mu}(\theta,
\varphi)h_{\ell',\mu'}(\theta, \varphi) = \delta_{\ell\ell'}\delta(\mu
- \mu'),
\label{ortho}
\end{equation}
\begin{equation}
\sum_{\ell=-\infty}^{+\infty}\!\!\int_{-\infty}^{+\infty}\!\!\!d\mu\;
h^*_{\ell,\mu}(\theta, \varphi)h_{\ell,\mu}(\theta', \varphi') =
\delta(\varphi - \varphi ')\frac{\delta(\theta -
  \theta')}{\sin\theta}.
\label{ortho1}
\end{equation}
Combining Eq.~(\ref{solution}) with Eq.~(\ref{ortho}) one obtains the
orthogonality of the electric modes
\begin{equation}
\int d^3\vec{r} \, \vec{E}^{\,*}{}^{\sigma}_{k,\ell,\mu}(\vec{r}) \cdot
\vec{E}^{\,\sigma'}_{k',\ell',\mu'}(\vec{r}) = \delta(k-k')\delta(\mu-\mu')
\delta_{\ell\ell'}\delta_{\sigma \sigma'}.
\label{orthoelectric}
\end{equation}

Let us consider an atomic qubit at a fixed position $\vec{r}$ in free
space with the transition dipole parallel to the $z$-axis.  Its
excited and ground state are denoted by $|e\rangle$ and $|g\rangle$
respectively.  The atom interacts with the quantized radiation field
distributed over a continuum of modes centered around the optical
atomic transition frequency $\omega_0$ and given by
Eq.~(\ref{solution}).  The standard dipole-interaction Hamiltonian of
such matter-field system reads $H \!= \!-\vec{d} \cdot
\hat{\vec{E}}(\vec{r})$ and simplifies for $\vec{d} = d \,\vec{e}_z$
to the following form
\begin{equation}
H_{int} \!=\! -i d\sqrt{\frac{\hbar c
    }{2\epsilon_0}}\sum_{\sigma,\ell}\int_0^{\infty}dk
  \sqrt{k}\int_{-\infty}^{+\infty}d\mu \left\{\vec{E}^{\,\sigma}_{k,\ell,\mu}(\vec{r})\vec{e}_z
  \hat{\sigma}^+\hat{a}^{\sigma}_{k,\ell, \mu} -\mathrm{h.c.}\right\},
\label{ham}
\end{equation}
where $\hat{\sigma}^+ \!\!=\!\! |e\rangle \langle g|$ and
$\hat{\sigma}^- \!\!=\!\!  |g\rangle \langle e|$ are the atomic rising
and lowering operators respectively.  This Hamiltonian shows that the
radiation field couples to the atom only if its polarisation has a
component which is parallel to the $z$-axis at the position of the
atom so that $\vec{e}^{\,\sigma}(\bar{n}) \cdot \vec{e}_z \not =0$.
From now on, we fix the frequency of the mode, $k=\omega_0/c$.  The spontaneous
emission decay rate for the atom immersed in the electromagnetic field
reservoir computed in the standard lowest order (Born) approximation
equals
\begin{eqnarray}
\Gamma(k,\vec{r}) \!\!&=& \!\!\frac{1}{(2 \pi)^2}\frac{d^2 k^3}{2\hbar
  \epsilon_0} \sum_{\sigma,\ell}\!\int_{-\infty}^{+\infty}\!\!d\mu
\int \!d\vec{n} \!\int \!d{\vec{n}}'
f_k(\vec{n},\vec{r})f^*_k(\vec{n}',\vec{r}) \nonumber\\ &{}& \cdot
e^{\sigma}_z(\vec{n}) e^{\sigma}_z(\vec{n}') \,
h^*_{\ell,\mu}(\vec{n},k) h_{\ell,\mu}(\vec{n}',k).
\label{gamma}
\end{eqnarray}
We denote here by $f_k(\vec{n},\vec{r})$ the plane wave $e^{i k\vec{n}
  \cdot \vec{r}}$ and $e^{\sigma}_z(\vec{n}) =
\vec{e}^{\,\sigma}(\vec{n}) \cdot \vec{e}_z$ and use the fact the the
summation over $\ell$ produces $\delta(\varphi -\varphi')$ (see
Eq.~(\ref{ortho1})). Therefore relevant $\vec{n}$ , $\vec{n}'$ and the
$z$-axis belong to the same plane what leads to the formula
\begin{eqnarray}
\sum_{\sigma} e^{\sigma}_z(\vec{n}) e^{\sigma}_z(\vec{n}') \!\!&=&\!\!
e^{2}_z(\vec{n}) e^{2}_z(\vec{n}') = \sin\theta \sin\theta'.
\end{eqnarray}
Taking into account that the integration over $\mu$ yields another
Dirac delta $\delta(\theta -\theta')$ we obtain
\begin{equation}
\Gamma(k;x,y,z) \!= \!\frac{1}{(2 \pi)^2}\frac{d^2 k^3}{2\hbar
  \epsilon_0}\int^{2 \pi}_0 \!\!\!d\varphi \!\!\int^{\pi}_0 \!\!d\theta
\sin^3\theta |f_k(x,y,z;\varphi,\theta)|^2,
\label{gamma1}
\end{equation}
where $\vec{r}\equiv (x,y,z)$ and $f_k(\vec{n},\vec{r})\equiv
f_k(x,y,z;\varphi,\theta)$. Knowing that
$|f_k(x,y,z;\varphi,\theta)|=1$, one recovers the standard result
\begin{equation}
\Gamma_0(k) \!=\! \frac{1}{3 \pi}\frac{d^2 k^3}{\hbar \epsilon_0}.
  \label{gammastandard}
\end{equation}

\section{Decay rate at the presence of a mirror}
\label{sec:mirror}

The presence of the conducting parabolic mirror leads to the boundary
conditions which should be imposed on the modes of the electric field
given by Eq.~(\ref{solution}).  Since it is very challenging to solve
the Helmholtz equation while keeping the zero-divergence condition
(Eq.~(\ref{trans})) and the boundary conditions satisfied at the same
time \cite{Nockel}, contrary to reference \cite{Nockel} we decided to
keep the transversality condition at the cost of relaxing the precise
value of the mode function on the mirror surface.  The transversality
condition relates the electric field to the geometry of the system and
therefore, contributes to some geometrical factor present in the decay
rate formula. The boundary condition ensures the discreteness of the
normal modes. However, the system is so large that the density of
modes can be approximated by a continuous spectrum. Enlarging the
system, thus changing the boundary conditions slightly, will not
influence the decay rate. Moreover, the boundary conditions are known
exactly only for stationary fields.

We first introduce the parabolic coordinates $(\xi, \eta, \varphi)$ related to Cartesian ones in the following way
\begin{eqnarray}
x &=& 2\sqrt{\xi \eta} \cos\varphi, \\ y &=& 2\sqrt{\xi \eta}
\sin\varphi, \\ z &=& \xi -\eta. 
\end{eqnarray}
The shape of the parabolic mirror is given by the equation
\begin{equation}
\eta = f.
\label{mirror}
\end{equation}
The normal modes (in fact their scalar counterparts) obtained by
separation method and expressed in parabolic coordinates are given by
products of the functions of $\xi , \eta, \varphi$, respectively
\cite{BKM}. We are interested in the $\eta $-dependent part
$F_{\ell,\mu}(k;\eta)$ which possesses the following asymptotic
behaviour
\begin{equation}
 F_{\ell,\mu}(k;\eta)\sim  \frac{\cos\left\{\mu\ln2k\eta +
      k\eta- \alpha_{\ell,\mu}\right\}}{\sqrt{\eta}},
  \label{asymptotic}
\end{equation}
where $\alpha_{\ell,\mu}$ is a certain phase. The boundary condition
imposed on Eq.~(\ref{asymptotic}) at the value $\eta = f$ can be
satisfied for a discrete set of $\mu_m$ only, which is related to the
periodicity of the cosine function.  We consider the simplest choice
\begin{equation}
kf-\alpha_{\ell,\mu}=0, \, \mu_m\ln 2kf = m\pi, \, m=1,2,3,...
\label{periodicity}
\end{equation}
This periodic condition is consistent with the replacement of the
continuous set of modes given by Eq.~(\ref{solution-free}) by a
discrete one
\begin{eqnarray}
\tilde{\chi}_{m}(\theta) &=& \frac{\sin\left(\frac{m\pi \ln[\tan
      \theta/2]}{\ln 2kf} \right)}{\sqrt{2\pi\ln
    2kf}\sin\theta}\;\mathrm{for} \; \theta\in [\theta_0 , \pi
  -\theta_0]\\ \nonumber &=& 0\ \mathrm{otherwise}
\label{solution-mirror}
\end{eqnarray}
such that
\begin{equation}
\tan\frac{\theta_0}{2}= \frac{1}{2kf}= \frac{1}{4\pi}\frac{\lambda}{f}.
\label{periodicity1}
\end{equation}
The limitation for the $\theta$ angle results from the quantization
condition and the fact that at the boundary the normal modes vanish
$\tilde{\chi}_{m}(\theta_0)=0$.  This anzatz modifies the formula for the
decay rate because the completeness condition now reads
\begin{equation}
\sum_m\tilde{\chi}_{m}(\theta)\tilde{\chi}_{m}(\theta')= I_{[\theta_0,
    \pi-\theta_0]}(\theta)\frac{\delta(\theta -\theta')}{\sin\theta},
\label{completeness}
\end{equation}
where $I_A$ denotes the indicator function of the set $A$. Therefore,
the integration over $\theta$ in Eq.~(\ref{gamma}) should be performed
over the interval $[\theta_0, \pi-\theta_0]$. This however, leads to
the correction of the order of $(kf)^{-4}$ which is completely
irrelevant from an experimental point of view. It is easy to notice
that while calculating the integral in Eq.~(\ref{gamma1}) in the
intervals $[0,\theta_0]$ and $[\pi-\theta_0,\pi]$.  In the experiment
prepared by the Erlangen group, the focal length is of order of
$f=2$mm and the wavelength of $\lambda=250$ nm, which amounts to
$kf\simeq 10^4$ \cite{Stobinska2009} and thus $\theta_0$ is
small. Therefore, we can replace $\sin\theta$ by $\theta$. It is
rather obvious that the same is true for any reasonable choice of the
boundary conditions, because the smallness of this correction is
entirely due to the large value of $kf$.

Hence, the only relevant modification of the spontaneous emission rate
due to the presence of the mirror is the replacement of the plane
traveling waves $f_k(\vec{n},\vec{r}) =e^{i k\vec{n} \cdot \vec{r}}$
by the standing waves
\begin{equation}
f_k(\vec{n},\vec{r}) =\sqrt{2}\sin\left( k\vec{n} \cdot
(\vec{r}-\vec{f})\right),
\label{standingwave}
\end{equation}
where the factor $\sqrt{2}$ ensures the completeness condition. Eq.~(\ref{standingwave}) 
implies that the electric field vanishes at the point P for any mode from Eq.~(\ref{solution}).  
It leads to the final expression for the spontaneous emission rate in the presence
of a conducting parabolic mirror
\begin{eqnarray}
&{}&\tilde{\Gamma}(k;x,y,z) = \frac{1}{2 \pi^2}\frac{d^2 k^3}{2\hbar
    \epsilon_0}\int^{2 \pi}_0 \!\!\!d\varphi \int^{\pi}_0 d\theta
  \sin^3\theta \nonumber\\ &{}& \sin^2\{k [(x \cos\varphi + y
    \sin\varphi)\sin\theta + (z+f) \cos\theta]\}.
\label{gamma2}
\end{eqnarray}

If the atom is placed at a distance from the point P much larger
than $\lambda$, the interference factor $\sin^2(...)$ is averaged to
$1/2$ and the standard free space result (Eq.~(\ref{gammastandard})) is
recovered.

\begin{figure}
\begin{center}
      \scalebox{0.6}{\includegraphics{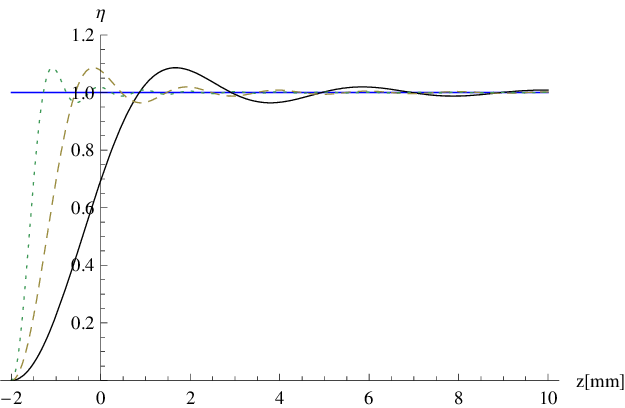}}
      \scalebox{0.6}{\includegraphics{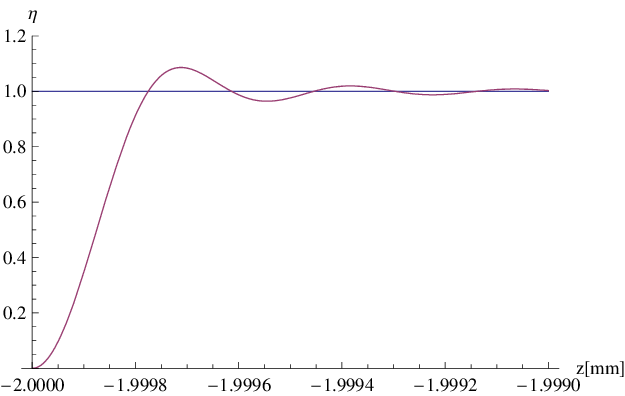}}
  \end{center}
  \caption{The spontaneous emission rate corrections evaluated for
    small values of the wavevector (the top figure): $k=0.25\pi$
    mm$^{-1}$ -- the solid line, $k=0.5\pi$ mm$^{-1}$ -- the dashed
    line, $k=\pi$ mm$^{-1}$ -- the dotted line and for large value of
    $k=10^4$ mm$^{-1}$ (the bottom figure).}
  \label{fig_corrections1}
\end{figure}

For the atom located at the mirror axis $x=y=0$, the integral in
Eq.~(\ref{gamma2}) simplifies to
\begin{equation}
\tilde{\Gamma}(k;z)= \eta \Gamma_0(k),
\end{equation}
where the correction to the free space decay rate is equal to
\begin{equation}
\eta = \left(1+3 \frac{\cos(2k(z+f))}{4k^2(z+f)^2} - 3 \frac{\sin(2
  k(z+f))}{8k^3(z+f)^3}\right).
\label{corrections}
\end{equation}
The correction $\eta$ is evaluated for $f=2$mm and depicted in
Fig.~\ref{fig_corrections1}. Its value at the focal point becomes
significant for small values of the wavevector $k$ (thus large values
of the wavelength $\lambda)$, corresponding to the condition
$|z+f|<\lambda$, i.e. for the atom which is close the the mirror
surface (within the distance of $\lambda$). However, if $k$ gets
large, then the $\eta$ fluctuations are shifted towards the mirror
surface and take place only at a length of the order of the
wavelength. Far away from the mirror all the fluctuations vanish,
$\eta=1$, and thus we observe a free-space decay rate, also at the
focus.  According to the figures, in the planned experiment
\cite{Stobinska2009,Markus} the changes in the decay rate could be
observable on a scale of $100$nm, but only within the distance of the
wavelength from the mirror surface.

\section{Conclusions}

We rigorously analyzed the modification of the electromagnetic vacuum
structure around an atom trapped at the focus of a parabolic metallic
mirror. We assumed that the atomic dipole moment is parallel to the
mirror axis. For a focal length which is large, compared to the
wavelength of the photon emitted during the atomic transition, the
total spontaneous emission rate does not differ from its
free-space value essentially.  However, the presence of the parabolic boundary
conditions may be revealed in a different way.  Since the energy
distribution among different modes is sensitive to the precise
position of the atom, one should observe interference effects on the
screen perpendicular to the mirror axis away from the focal
point. They are analogous to those observed in the experiment with an
atom trapped in front of a flat mirror~\cite{Eschner2001}. The dipole
radiation has to obey the boundary condition: the field has to vanish
on the mirror surface, only those modes will contribute to the pattern
on the screen which fulfill this condition. The other modes will be
suppressed and will give rise to dark fringes. The detailed structure
of the fringes depends on the value $kf$ and the precise position of
the atom.

\section*{Acknowledgments}

This work was supported by the EU 7FP Marie Curie Career Integration
Grant No. 322150 ``QCAT'', MNiSW grant No. 2012/04/M/ST2/00789 and FNP
Homing Plus project.  R.A. is supported by the Polish research network
LFPPI.  The authors thank G. Leuchs, L.~L.~S\'anchez-Soto and G. Alber
for stimulating discussions.

\end{document}